\documentclass[prl,aps,
twocolumn,
preprintnumbers,
nofootinbib,
floatfix,
superscriptaddress]{revtex4}
\usepackage{float}

\usepackage{amsmath,amssymb}
\usepackage{graphicx}
\usepackage{feynmf}
\usepackage{xcolor}
\usepackage{hyperref}
\definecolor{darkpurple}{rgb}{0.5, 0.2, 0.8}
\definecolor{darkblue}{rgb}{0.0, 0.0, 0.8}
\definecolor{darkgreen}{rgb}{0.0, 0.4, 0.0}
\definecolor{darkred}{rgb}{0.5, 0.0, 0.0}
\hypersetup{
    linktocpage,
     colorlinks,
     citecolor=darkgreen,
     linkcolor= darkgreen,
     urlcolor=darkgreen
}

\newcommand{\Li}{ {\mathrm{Li}}}

\begin{document}
\setlength{\parskip}{0pt}
\begin{fmffile}{feyngraph}

\title{Implications of the Landau Equations for Iterated Integrals}

\author{Holmfridur S. Hannesdottir}
\affiliation{Department of Physics, Harvard University, Cambridge, MA 02138, USA}
\affiliation{Institute for Advanced Study, Einstein Drive, Princeton, NJ 08540, USA}
\author{Andrew~J.~McLeod}
\affiliation{Niels Bohr International Academy and Discovery Center, Niels Bohr Institute\\
University of Copenhagen, Blegdamsvej 17, DK-2100, Copenhagen \O, Denmark}
\author{Matthew D. Schwartz}
\affiliation{Department of Physics, Harvard University, Cambridge, MA 02138, USA}
\author{Cristian~Vergu}
\affiliation{Niels Bohr International Academy and Discovery Center, Niels Bohr Institute\\
University of Copenhagen, Blegdamsvej 17, DK-2100, Copenhagen \O, Denmark}

\begin{abstract}
We introduce a method for deriving constraints on the symbol of Feynman integrals from the form of their asymptotic expansions in the neighborhood of Landau loci. In particular, we show that the
behavior of these integrals near singular points is directly related to the position in the symbol where one of the letters vanishes or becomes infinite. We illustrate this method on integrals with generic masses, and as a corollary prove the conjectured bound of $\lfloor \frac {D \ell} 2\rfloor$ on the transcendental weight of polylogarithmic $\ell$-loop integrals of this type in integer numbers of dimensions $D$. We also derive new constraints on the kinematic dependence of certain products of symbol letters that remain finite near singular points.
\end{abstract}

\maketitle

The analytic structure of scattering amplitudes is strongly constrained by physical principles such as locality and causality. These principles imply, for instance, that amplitudes can only develop branch cuts at solutions to the Landau equations---where propagators are put on shell and particles interact in physically possible configurations~\cite{landau1959}---and that double discontinuities in partially-overlapping momentum channels must vanish~\cite{Steinmann,Steinmann2,Cahill:1973qp}. 
In this paper, we 
derive new constraints on the branch cut structure of Feynman integrals from considerations of their behavior near Landau loci. We do this by connecting the leading non-analytic behavior of a generic polylogarithmic expression near one of its singular points
to its various integral representations.
We also identify new constraints on the logarithmic branch points that appear next to these singularities in iterated integrals
that follow from integrability.
To illustrate our approach, we focus in this Letter on constraints for Feynman integrals involving generic masses, whose leading non-analytic behavior near singular points was worked out by Landau~\cite{landau1959}. Even for this class of integrals,
the methods we introduce have powerful implications, such as
bounding the number of integrations that can appear when they are expressed in terms of iterated integrals.

In carrying out our analysis, we are assisted by the symbol map~\cite{Goncharov:2010jf}, which encodes how polylogarithms can be written as linear combinations of iterated integrals of $d \log$ differential forms. We thus build on a growing body of literature that has leveraged the natural connection between the study of symbols of Feynman integrals and Landau analysis~\cite{Abreu:2014cla,Dennen:2015bet,Dennen:2016mdk,Abreu:2017ptx,Prlina:2017azl,Prlina:2017tvx,Prlina:2018ukf,Gurdogan:2020tip}. More generally, the symbol has been observed to encode many important features of the analytic structure of Feynman integrals, such as the Steinmann relations~\cite{Steinmann,Steinmann2,Cahill:1973qp,Caron-Huot:2016owq,Bourjaily:2020wvq}, and the more extensive cluster adjacency conditions (or extended Steinmann relations) that have been observed in planar $\mathcal{N} = 4$ supersymmetric Yang-Mills theory~\cite{Drummond:2017ssj,Caron-Huot:2019bsq}. 

In~\cite{landau1959}, Landau first identified the set of kinematic loci where a Feynman integral can become singular (see~\cite{ELOP} for a pedagogical presentation). 
To do so, he considered a generic $\ell$-loop Feynman integral with $n$ propagators in $D$ dimensions, which takes the form
\begin{equation}
    \label{eq:Feynman_integral}
    I(p_i) = \int \frac{d^D k_1\cdots d^D k_\ell}{A_1 \cdots A_n} \, ,
\end{equation}
where the propagators $A_i$ are quadratic in the external momenta $p_i$ and loop momenta $k_i$. When combined with Feynman parameters $\alpha_i$, this integral can be rewritten as
\begin{equation}
 I(p_i) = (n-1)! \int \frac{d^{D \ell} k \; d^n \alpha}{M^n}
    \delta\Big(1 - \sum_{i=1}^n \alpha_i\Big) \, ,
\end{equation}
where
\begin{equation}
    M = \alpha_1 A_1 + \cdots + \alpha_n A_n \, .
\end{equation}
In this form, we see that $I(p_i)$ will only be singular where $M$ vanishes, if either $\alpha_i=0$ or $A_i=0$ for each $i$.

The leading singularity of $I(p_i)$ will be given by solutions to the $M=0$ constraint in which all $\alpha_i \ne 0$, while subleading singularities correspond to solutions in which a subset of these Feynman parameters vanish. 
However, these conditions are not yet sufficient to guarantee that $I(p_i)$ is singular; if these conditions are satisfied at a generic point in the integration region, the integration contour can be deformed to avoid the corresponding singularity. To encounter an actual singularity, the contour must be pinched: at least two zeroes of the denominator must coalesce, which is equivalent to requiring both that $M=0$ and $\frac{\partial}{\partial k_i} M = 0$. Thus, the Landau equations---which completely characterize the singular surface, called the {\emph{Landau locus}}---are given by $\frac{\partial}{\partial k_i} M = 0$ and either $\smash{\frac{\partial}{\partial \alpha_i} M = 0}$ or $\alpha_i = 0$ for all $i$. For now, we focus on the leading singularity, and impose the conditions that $\smash{\frac{\partial}{\partial \alpha_i} M = 0}$ and $\smash{\frac{\partial}{\partial k_i} M = 0}$ for all $i$.

In the same paper~\cite{landau1959}, Landau also characterized the nature of the singularity in a {\emph{Landau limit}}, as the Landau locus is approached. To do this, one first completes the square and shifts $k_i \to k_i'$ to remove all terms in the denominator that are linear in the loop momenta. This leads to 
\begin{equation}
    M=\varphi(p_i, \alpha_i) + K(k_i',\alpha_i)  \, ,
\end{equation}
where $K$ is a homogeneous quadratic form in the shifted loop momenta $k_i'$.
After this change of variables, the Landau equations become $\varphi=0$ and $K=0$ at the pinched singular surface, in addition to $\smash{\frac{\partial}{\partial k_i'} K =0}$ for all $i$. 

Let us now consider a phase space point $\bar{p}_{i}$ close to this singular surface, and let $\bar{\alpha}_{i}$ be the (nonzero) values of $\alpha_i$ that minimize $\varphi$ at $\bar{p}_{i}$. Writing $\alpha_i = \bar{\alpha}_{i} + \alpha_i'$, we then have
\begin{equation} \label{eq:D_expansion}
    M = \varphi_0 + Q(k_i',\alpha_i')+ \cdots \, ,
\end{equation}
where $\varphi_0 = \varphi(\bar{p}_i,\bar{\alpha}_{i})$ and $Q$ is now quadratic in the Feynman parameters $\alpha_i'$ as well as the loop variables $k_i'$. All remaining contributions, which we have omitted, are cubic or higher in the the $k_i'$ and $\alpha_i'$ variables. 

The nature of the singularity near $\varphi_0$ is determined by the behavior of the integral in the region where the value of $M$ is close to $\varphi_0$. Thus, it is sufficient to consider values of $k_i'$ and $\alpha_i'$ that are below some small cutoff $\delta$. Collecting the $\ell D$ loop momenta and $n-1$ Feynman parameters into a vector $\vec{x}$ of dimension $m=\ell D+n-1$, the nature of the singularity is characterized by the integral
\begin{equation} \label{eq:integral_near_singular_surface}
    I(\varphi_0) \sim \int_0^\delta \frac{d^m x}{(\varphi_0 + \vec{x}^2)^n} \sim \varphi_0^\gamma \int_0^{{\delta}/{\sqrt{\varphi_0}}} \frac{r^{m-1} dr}{(1+r^2)^n} \, ,
\end{equation}
where we have rescaled $\vec{x}\to \sqrt{\varphi_0}\vec{x}$ and gone to spherical coordinates in the second step.
We call $\gamma=\frac{m}{2}-n=\frac{1}{2}(\ell D - n -1)$ the {\emph{Landau exponent}}. If $\gamma<0$, the integral multiplying $\varphi_0^\gamma$ is convergent as $\varphi_0 \to 0$ and then the singular behavior is $I(\varphi_0)\sim \varphi_0^\gamma$. If $\gamma>0$, then this integral diverges as $\varphi_0 \to 0$. In this case, one can first take $\lceil{\gamma}\rceil$ derivatives of $I(\varphi_0)$, evaluate the integral, and then integrate with respect to $\varphi_0$ the same number of times. The result is that the dominant singular behavior is $I(\varphi_0) \sim \varphi_0^\gamma \log \varphi_0$ if $\gamma$ is an integer and $I(\varphi_0) \sim \varphi_0^\gamma$ if $\gamma$ is half-integer. Finally, if $\gamma=0$, the integral gives $I(\varphi_0)\sim \log \varphi_0$. Thus, Landau's result for the nature of the singularity is that 
\begin{equation} \label{eq:landau_singularity_nature}
    I(\varphi_0) \sim \begin{cases}
   C  \varphi_0^\gamma \log \varphi_0   & \text{if } \gamma \in \mathbb{Z}, \gamma \ge 0 \\ 
   C \varphi_0^\gamma & \text{otherwise.}
    \end{cases}
\end{equation}
Alternatively, one can simply evaluate $I(\varphi_0)$ directly in terms of hypergeometric functions, and then expand at small $\varphi_0$ to deduce the singular behavior. A formula for the prefactor $C$ in terms of the Hessian $H$ of $M$ with respect to $\alpha_1, \dotsc, \alpha_{n-1},$ evaluated at the solution of the Landau equations, can be found in~\cite{PolkinghorneScreaton}, where it is shown that $\smash{C \propto (\det H)^{-\frac{1}{2}}}$. 

The same analysis can also be carried out for subleading singularities, where some of the $\alpha_i$ are expanded around zero. The result for the Landau exponent $\gamma$ is the same as above, with $n$ reinterpreted as the number of nonzero Feynman parameters rather than the number of internal lines in the original diagram, and $L$ reinterpreted as the number of loops in the diagram after the edges with vanishing Feynman parameters have been contracted out~\cite{ELOP,PolkinghorneScreaton}.

Some comments are in order.
First, note that the Landau singularities arise from the integration region where the loop momenta $k'$ are small. Although the logarithmic behavior arises from the region where $\smash{\frac{k'}{\sqrt{\varphi_0}}\to \infty}$, this does not require the loop momenta $k'$ to take values larger than $\delta$.
If the Feynman integral is infrared divergent, as may happen when massless particles are involved, multiple singularities can converge and the analysis is more complicated. Even so, we note that the asymptotic expansion near a Landau locus can be computed exactly in the dimensional regularization parameter $\epsilon$ using the results of~\cite{PolkinghorneScreaton}. We leave the more involved analysis of this expansion in dimensional regularization to future work.

Second, we highlight the fact that the singular behavior characterized in~\eqref{eq:landau_singularity_nature} is the leading non-analytic behavior, but not necessarily the dominant behavior in the expansion around $\varphi_0=0$. For example, when $\gamma=1$, we will generically find $I(\varphi_0) \sim \varphi_0 \log \varphi_0 + \text{const}$ for small $\varphi_0$. 

Third, Landau's result in~\eqref{eq:landau_singularity_nature} must be generalized if some of the integration variables do not appear at quadratic order in the expansion near the singular locus in~\eqref{eq:D_expansion}, or equivalently if the determinant of the Hessian $H$ is zero. This may happen when the the internal masses are not generic. For example, 
the bowtie graph with reflection symmetry in the masses has a double-logarithmic singularity:
\begin{equation}
\begin{gathered}
{\includegraphics[trim=0 0 13cm 0,clip,width=0.2\textwidth]{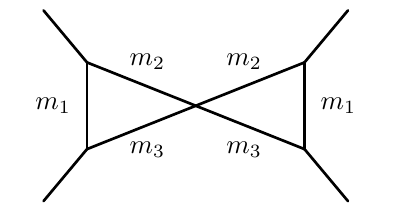}}
\end{gathered}
\sim \log^2\varphi_0
\end{equation}
where $\varphi_0$ matches the quantity that vanishes in one of the singular limits of a single triangle integral, whose Landau equations are solved in~\cite{Eden:2011yp}. Such factorized integrals arise when describing subleading singularities of non-factorizing integrals with specific internal masses (for example, the two-loop ladder collapses to the bowtie when the middle rung is contracted).

Finally, Landau's result must also be revised if $\varphi_0$ is identically zero. Then, one should consider the expansion around the first non-vanishing term in the denominator rather than around $\varphi_0$. This type of situation arises, for instance, for subleading singularities in the sunrise integral. In this work, we restrict to situations in which $\varphi_0 \neq 0$ and $\det H \neq 0$.

With these subtleties out of the way, let us return to the types of singularity that arise in Eq.~\eqref{eq:integral_near_singular_surface} for integer $\gamma$. As $\varphi_0$ approaches zero with real on-shell momenta (on the physical sheet), $\varphi_0^\gamma \log \varphi_0$ is non-singular for $\gamma >0$. However, Feynman integrals typically evaluate to functions whose maximal analytic continuation away from the physical sheet strongly constrains their possible singular behavior. For example, the classical polylogarithm $\Li_{\gamma+1}(1-\varphi_0)$ has a branch point at $\varphi_0=0$ near which its leading non-analytic behavior takes the form seen above, $\Li_{\gamma+1} (1-\varphi_0) \sim \varphi_0^{\gamma} \log \varphi_0$. Although $\lim_{\varphi_0 \to 0} \varphi_0^{\gamma} \log \varphi_0 =0$ for $\gamma >0$, 
one can encircle the branch point at $\varphi_0 = 0$ and find additional singularities on higher Riemann sheets. Conversely, if one knows that a function is a classical polylogarithm, and is told that the leading non-analytic behavior near $\varphi_0 = 0$ is $\varphi_0^\gamma \log \varphi_0$ on the principal branch, one can immediately deduce this function must be $\Li_{\gamma+1}(1-\varphi_0)$. 
Classical polylogarithms are the simplest examples of the types of iterated integrals that appear ubiquitously in Feynman integrals. Thus, we would similarly like to know how such iterated integrals are constrained by their known form in Landau limits. 

We thus consider an iterated integral $F(s_i)$ that depends on some arguments $s_i$ (internal masses and Lorentz-invariant combinations of external momenta). We take one of the independent arguments in $s_i$ to be $\varphi_0$, and study the limit $\varphi_0 \to 0$.
For simplicity, we assume that the function $F(s_i)$ is polylogarithmic and thus has a symbol~\cite{Goncharov:2010jf}, which can be written as
\begin{equation} \label{eq:F_symbol}
\mathcal{S}\Big[ F(s_i) \Big] = \sum a_1(s_i) \otimes \cdots \otimes a_n(s_i) \, ,
\end{equation}
where we have left the sum over terms in the symbol schematic.\footnote{We work at the level of the symbol throughout, as this will be sufficient for establishing our main results. However, it is straightforward to generalize the argument in this section to include contributions involving transcendental constants.} Equation~\eqref{eq:F_symbol} represents the fact that $F(s_i)$ can be expressed as a sum of iterated integrals over the forms $d\log a_j (s_i)$, where the integration contour begins at some chosen basepoint $s_{i}^\bullet$ and ends at the point $s_i$. 

We can formulate the iterated integrals represented by these terms in the symbol explicitly by pulling back the integral over each $d\log a_j (s_i)$ to an auxiliary space of variables $t_j$, in which the integration contour is given by $0\le t_1\le t_2 \le \cdots \le t_n \le 1$. For this purpose, it is simplest to choose the straight path
\begin{equation}
   \sigma_i(t) = (1-t) s_{i}^\bullet + t s_i \, .
\end{equation}
Then, the pullbacks are given by
\begin{equation}
  \sigma^* (d \log a_j)(t) =  \frac {(s_i - s_{i}^{\bullet}) \cdot (\nabla_i a_j)(\sigma(t)) }{a_j(\sigma(t))} d t \, .
\end{equation}
For example, if we take the basepoint in the $\varphi_0$ variable to be $\varphi_0^\bullet = 1$, the pullback along a straight path from the basepoint to a generic value of the arguments $s_i$ (along which the other variables are also allowed to vary) will be given by 
\begin{equation} 
  \sigma^* (d \log \varphi_0)(t) = \frac {\varphi_0 - 1}{1 - t + t \varphi_0} dt \, .
  \label{phipullback}
\end{equation}
After changing variables using the pullback, the symbol in~\eqref{eq:F_symbol} can be rewritten in terms of iterated integrals as
\begin{multline} \label{eq:F_iterated_integral}
    F(s_i) = \sum \int_{0\leq t_1 \leq \cdots \leq t_n \leq 1} \sigma^*(d \log a_1)(t_1)   \\ 
    \times \sigma^*(d \log a_2)(t_2) \cdots 
    \sigma^*(d \log a_n)(t_n) \, .
\end{multline}
Importantly, all of the dependence on the variables $s_i$ is now in the integrands $\sigma^* (d \log a_j(s_i)) (t)$. 

Let us now study the $\varphi_0 \to 0$ limit of $F(s_i)$. We do this by expanding each of the forms $d \log a_j(s_i)$ within the integrand of Eq.~\eqref{eq:F_iterated_integral} near the $\varphi_0 = 0$ hypersurface. Individual $d \log$s may become singular as they approach this hypersurface, if the corresponding symbol letter $a_j(s_i)$ approaches zero or infinity. In generic polylogarithms, any number of symbol letters may become singular in each term of the symbol. We will here work out the case in which at most a single letter becomes singular, as this will allow us to connect to the leading non-analytic behavior Feynman integrals with generic masses are expected to have from  Eq.~\eqref{eq:landau_singularity_nature}.\footnote{Even in these integrals, it may be that multiple letters become singular in individual terms of the symbol; however, Landau's argument tells us these contributions must be subleading.} Symbol terms in which multiple letters become singular can be analyzed using the same approach.

Thus, let us analyze the contribution to $F(s_i)$ coming from a term in the symbol that involves a single letter $\varphi_0$ in the limit $\varphi_0 \to 0$. Such a term generically takes the form
\begin{equation} \label{eq:generic_S1_term}
b_1 \otimes \cdots \otimes b_p \otimes \varphi_0 \otimes c_1 \otimes \cdots \otimes c_q
\end{equation}
for some $p, q \ge 0$, where none of the $b_i$ or $c_i$ depend on $\varphi_0$ at leading order (and we have left their dependence on the rest of the variables $s_i$ implicit). As an iterated integral, the contribution from this term can be written as
\begin{equation} \label{eq:term_contribution}
f(\varphi_0) = \int_0^1 U(t) \, \sigma^*(d \log \varphi_0)(t) \, V(t) \, ,
\end{equation}
where
\begin{equation}
U(t) = \int_0^t \sigma^*(d \log b_1)(t_1) \cdots \int_{t_{p - 1}}^t \!\!\sigma^*(d \log b_p)(t_p) \,  ,
\end{equation}
and
\begin{equation}
V(t) = \int_t^1 \sigma^*(d \log c_1)(t_1) \cdots \int_{t_{q - 1}}^1 \!\! \sigma^*(d \log c_q)(t_q) \, ,
\end{equation}
and the pullback of $d \log \varphi_0$ is given in Eq.~\eqref{phipullback}.

As can be seen in Eq.~\eqref{phipullback}, the logarithmic singularity arises in Eq.~\eqref{eq:term_contribution} near the $t = 1$ boundary. As $t\to 1$, the function $U(t)$ will generically limit to a nonzero value (independent of $\varphi_0$)
while $V(t)$ will vanish, since its integration interval shrinks to zero. Thus, to determine the leading contribution to $f(\varphi_0)$, we consider the expansion of $V(t)$ near $t=1$. This requires computing derivatives of $V(t)$ with respect to $t$; for instance, 
\begin{multline}
  \frac d {d t} V(t) = -\frac {\sigma^*(d \log c_1)}{d t}(t) \\ 
  \times\int_t^1 \sigma^*(d \log c_2)(t_2) \cdots \int_{t_{q - 1}}^1 \sigma^*(d \log c_q)(t_q) \, .
\end{multline}
When $q \geq 2$, the integration interval in the remaining integrals will still shrink to zero, so $\frac d {d t} V(1)$ will also vanish. The first nonzero derivative of $V(t)$ at $t = 1$ will thus be
\begin{equation}
  \frac {d^q V} {d t^q}(1) = (-1)^q \frac {\sigma^*(d \log c_1)}{d t}(1) \cdots \frac {\sigma^*(d \log c_q)}{d t}(1)\, ,
\end{equation}
where we have dropped all terms in which multiple derivatives act on a single logarithm, since these terms also vanish. Integrating this constant in the region of $t=1$ gives us the leading contribution to Eq.~\eqref{eq:term_contribution}, namely
\begin{equation} \label{eq:leading_order_integral}
f(\varphi_0) \sim  U(1) \frac 1 {q!} \frac {d^q V}{d t^q}(1) \int_0^1 (t - 1)^q \frac {(\varphi_0-1)d t}{1 - t + t \varphi_0},
\end{equation}
where we have plugged in the explicit form of $\sigma^* (d \log \varphi_0)(t)$ from  Eq.~\eqref{phipullback}.

We can evaluate Eq.~\eqref{eq:leading_order_integral} using the change of variables $u = 1 - t + t \varphi_0$, which allows us to rewrite
\begin{equation}
  \int_0^1 (t - 1)^q \frac {(\varphi_0 - 1) d t}{1 - t + t \varphi_0} = -\frac  1 {(\varphi_0-1)^q} \int_{\varphi_0}^1 \frac {d u} u (u - \varphi_0)^q.
\end{equation}
This form of the integral can be computed using the binomial expansion of $(u-\varphi_0)^{q}$. All the terms in this expansion will be polynomials in $u$ that can be integrated rationally, except the term $(-\varphi_0)^q \frac {d u} u$. Since we are only interested in the non-analytic contributions, we drop all the polynomial terms and obtain 
\begin{equation} \label{eq:leading_nonholomorphic_symbol_term}
   f(\varphi_0) \sim
   U_p(1) \frac 1 {q!} \frac {d^q V_q}{d t^q}(1) \varphi_0^q \log \varphi_0.
\end{equation}
This is the leading non-analytic behavior contributed to $F(s_i)$ by each symbol term of the form in Eq.~\eqref{eq:generic_S1_term}.

By connecting this result to the leading non-analytic behavior of a Feynman integral in one of its Landau limits, we can place a bound on how close to the end of the symbol the corresponding branch point can appear. For instance, for Feynman integrals with generic masses, the contributions in Eq.~\eqref{eq:leading_nonholomorphic_symbol_term} take the form predicted by Landau if we identify $q = \gamma$ and $\smash{C = \frac {1} {q!} U_p(1) \frac {d^q V_q}{d t^q}(1)}$. 
 Eq.~\eqref{eq:leading_nonholomorphic_symbol_term} therefore implies that 
the leading non-analytic contributions to $F(s_i)$ come from terms with the smallest $q$, namely those in which $\varphi_0$ appears closest to the end of the symbol.

This result further allows us to derive a bound on the transcendental weight of $\ell$-loop Feynman integrals with generic masses in $D$ dimensions (that is, the number of letters appearing in each term of their symbol).\footnote{Transcendental weight has also been extended to elliptic polylogarithms~\cite{Broedel:2017kkb,Broedel:2018iwv,Broedel:2018qkq}, and may generalize to more complicated classes of integrals appearing in Feynman diagrams~\cite{Brown:2009ta,Brown:2010bw,Bloch:2014qca,Bloch:2016izu,Bourjaily:2018ycu,Bourjaily:2018yfy,Bourjaily:2019hmc}.} Eq.~\eqref{eq:leading_nonholomorphic_symbol_term} implies that the Landau limits with the largest integer Landau exponent $\gamma$ describe the logarithmic singularities that can appear furthest from the end of the symbol. Recalling that $\gamma = \frac 1 2 (D \ell - n - 1)$ from Landau's analysis, we see the largest $\gamma$ occurs when the number of nonzero Feynman parameters $n$ is minimized. Requiring that $\gamma$ is an integer (so the singularity is logarithmic) and that at least one Feynman parameter is nonzero, we get 
$\gamma \le \lfloor \frac {D \ell} 2\rfloor -1$. 
The transcendental weight of an $\ell$-loop Feynman integral with generic masses in $D$ dimensions can be at most one larger than this maximum integer Landau exponent, and is thus bounded by $\lfloor \frac {D \ell} 2\rfloor$.\footnote{The vanishing of a symbol entry cannot necessarily be achieved with
momenta in the physical region. There is nevertheless a correspondence with the singular
behavior from Landau's analysis if one continues to complex momenta and complex values of $\alpha_i$.}

Note that the maximum transcendental weight of these integrals increases differently with the loop order in even and odd numbers of dimensions. In even dimensions, the maximum weight increases by $\frac{D}{2}$ with each additional loop, while in odd dimensions the maximum transcendental weight increments alternately by $\frac{D-1}{2}$ or $\frac{D+1}{2}$ as $\ell$ increases.  For example, when $D = 3$, the maximum weight is $1$ at one loop, at two loops it is $3$, at three loops it is $4$, and so on.

It may also seem curious that the logarithmic singularity that appears in the first entry of the symbol in even dimensions is a subleading Landau singularity in which a single Feynman parameter is nonzero. In particular, one might have expected that these first entries would correspond to bubble Landau diagrams. However, one can see from Landau's analysis that these bubble diagrams correspond to square root singularities in this class of Feynman integrals. 

We emphasize that we can use the result in Eq.~\eqref{eq:leading_nonholomorphic_symbol_term} to predict the position in the symbol at which the logarithmic singularities associated with specific Landau diagrams are expected to appear, by computing the associated Landau exponent.
While this procedure only identifies the locations at which logarithmic singularities can appear at various positions in the symbol, and not the symbol letters themselves, it still provides a similar restriction to the first and last entry conditions that restrict the letters that can appear in certain positions in the symbol in planar $\mathcal{N}=4$ supersymmetric Yang-Mills theory~\cite{Gaiotto:2011dt,Caron-Huot:2011zgw,Caron-Huot:2011dec,Dixon:2015iva,He:2019jee,He:2020vob,He:2021mme}. These types of conditions have played a crucial role in bootstrapping supersymmetric amplitudes~\cite{Caron-Huot:2020bkp}, so the fact that similar predictions can be made for individual Feynman integrals at all positions in the symbol using Landau analysis is highly encouraging.

In fact, it is possible to check that  the Landau exponent correctly predicts the position of the logarithmic singularities that appear in the case of the one-loop all-mass integrals studied in~\cite{aomoto1977,Bourjaily:2019exo}. There, the $D$-gon integral is considered in $D$ dimensions, and its symbol can be computed using the Schl\"afli formula~\cite{Schlaefli:1860}. This symbol has
transcendental weight $\lfloor \frac {D} 2\rfloor$ (thus saturating the expected bound), and the codimension-one limits in which logarithmic singularities can appear at each position in the symbol can be explicitly identified. In particular, even $D$-gons have logarithmic singularities associated with all odd $q$-gon Landau diagrams for $q < D$, while odd $D$-gons have logarithmic singularities associated with all even $q$-gon Landau diagrams for $q < D$. These subleading singularities all appear at the position in the symbol predicted by the Landau exponent. More details can be found in~\cite{to_appear}. 

Finally, let us consider the implications of integrability for the symbols of Feynman integrals near a Landau locus. In a codimension-one Landau limit described by $\varphi_0 \to 0$, a weight-two symbol can be put in the form
\begin{equation} \label{eq:generic_weight_two_limit}
\sum_i \varphi_0 \otimes a_i + \sum_j b_j \otimes \varphi_0 + \sum_{k, l} c_k \otimes d_l \, ,
\end{equation}
where we have pulled the coefficients in front of the symbol terms into the exponents of the letters $a_i$, $b_j$, $c_k$, and $d_l$. At order $\varphi_0^{-1}$ integrability implies
\begin{equation}
\label{int}
\sum_i \frac {d \varphi_0} {\varphi_0} \wedge \frac {d a_i}{a_i} + \sum_j  \frac {d b_j}{b_j} \wedge \frac {d \varphi_0} {\varphi_0}   = 0 \, ,
\end{equation}
which will only be satisfied if
\begin{equation} \label{eq:constant_ratio}
\lim_{\varphi_0 \to 0} \frac{\prod_i a_i}{\prod_j b_j} = \text{constant}\, 
\end{equation}
for some nonvanishing constant that is independent of all kinematics. 
Further relations are imposed by integrability at higher orders in $\varphi_0$, when the letters $a_i$, $b_j$, $c_k$ and $d_l$ are expanded beyond leading order. We will not write these relations down in detail. All of these identities must also hold in adjacent entries of symbols of higher weight.

In the simplest cases, when the symbol in Eq.~\eqref{eq:generic_weight_two_limit} is just that of a product of logarithms, the constraint in Eq.~\eqref{eq:constant_ratio} will be satisfied due to the fact that the product of letters in the numerator and denominator will be identical. However, one can more generally encounter examples in which these sets of letters are different; in fact, it will often be the case that the product over $i$ or $j$ is empty. For instance, we have checked various Landau limits of the all-mass box, hexagon, and heptagon symbols from~\cite{aomoto1977,Bourjaily:2019exo}, where the product over either $i$ or $j$
is empty, and have observed that the remaining product becomes 1 as $\varphi_0 \to 0$. In such cases, equation~\eqref{eq:constant_ratio} provides a surprising and nontrivial constraint. 

In this paper, we have introduced new methods for deriving constraints on the symbol of Feynman integrals. We have illustrated this method with the example of Feynman integrals with generic masses and trivial numerators, as these integrals fall within the scope of Landau's analysis and their leading non-analytic behavior in Landau limits is known.\footnote{Nontrivial numerators generically do not affect the Landau equations~\cite{Collins:2020euz}, and
only reduce the degree of divergence in singular limits. Indeed, in certain approaches to computing amplitude integrands, numerators are specifically chosen to vanish on the support of certain Landau loci (see for instance~\cite{Campbell:1996zw,Denner:2005nn,Berger:2008sj,Bourjaily:2017wjl,Bourjaily:2019iqr,Bourjaily:2019gqu,Bourjaily:2020qca}).} This implies, in particular, that they involve only a single power of $\log \varphi_0$ near codimension-one Landau limits.
We have also restricted our attention to polylogarithmic iterated integrals. However, we note that these results also apply to Feynman integrals that have subtopologies involving generic masses. The behavior of such integrals near Landau limits that involve putting only these propagators with generic masses on shell will also be described by Eq.~\eqref{eq:landau_singularity_nature}, allowing similar constraints to be derived on the position of the corresponding branch points in their symbols.

While a similar analysis of the behavior of more general Feynman integrals near Landau loci  will be more complicated, we anticipate the same strategy can be used to derive analogous results. 
For instance, it should be straightforward to generalize to 
iterated integrals that are not of $d \log$ forms,
such as those containing elliptic integration kernels.
Further investigation into Feynman integrals with massless particles, non-integer dimensions, and behavior involving higher powers of $\log \varphi_0$ in singular limits could lead to additional insight into the structure of Feynman integrals. It should be possible to derive results analogous to Eq.~\eqref{eq:landau_singularity_nature} in each of these cases using the results of~\cite{PolkinghorneScreaton}, which would allow constraints to be derived on the symbols of these integrals using Eq.~\eqref{eq:leading_nonholomorphic_symbol_term} (or using the generalization of this equation to iterated integrals involving kernels beyond $d \log$s). 

Recent advances in our understanding of the analytic structure of scattering amplitudes have greatly facilitated the computation of nontrivial amplitudes to high loop orders. 
This has been most striking in bootstrap computations,
where knowledge about the analytic structure of amplitudes provides important constraints (see for instance~\cite{Dixon:2016nkn,Drummond:2018caf,Caron-Huot:2019vjl,Dixon:2020bbt}). The additional structure that we have observed here regarding the positions of logarithmic singularities in the symbol of Feynman integrals and nontrivial restrictions among products of symbol letters near Landau limits should amplify the power of bootstrap methods, applied both to amplitudes and individual Feynman integrals (as done in~\cite{Chicherin:2017dob,Henn:2018cdp,Caron-Huot:2018dsv,He:2020lcu,He:2021non,He:2021esx}). In particular, our results should help extend bootstrap methods to amplitudes and Feynman integrals that involve internal masses. 
For example, the constraints derived here, in combination with integrability, may allow one to predict the symbol letters that appear in the one-loop all-mass integrals studied in~\cite{Bourjaily:2019exo}. With the
help of the hierarchical principle~\cite{Landshoff1966}, one could then consider bootstrapping these integrals and possibly even higher-loop all-mass integrals, which have proven extremely resistant to direct computation.

\vspace{.3cm}

\paragraph{Acknowledgments}  We are grateful to Jacob Bourjaily and Lance Dixon for the initial collaboration out of which this project developed.  We also thank Nima Arkani-Hamed, Hjalte Frellesvig, Enrico Herrmann, Matt von Hippel, Sebastian Mizera, and Matthias Volk for discussions. This work has been supported in part by the U.S.\ Department of Energy under contract DE-SC0013607 (MS,HSH), and an ERC Starting Grant (No.\ 757978) and grant from the Villum Fonden (No.\ 15369) (AJM,CV), and a grant from the Simons Foundation (816048, HSH).
\end{fmffile}

\bibliographystyle{physics}
\bibliography{asymptotics}

\end{document}